\documentclass[12pt]{iopart}

\usepackage{iopams}  

\usepackage{graphics} 
\usepackage{graphicx}  
\usepackage{amsfonts}  
\usepackage{amssymb}
\usepackage{mathrsfs}
\usepackage{subfigure}
\usepackage{color}

\renewcommand{\phi}{\varphi}

\newcommand{\be}{\begin{equation}}
\newcommand{\ee}{\end{equation}}
\newcommand{\bea}{\begin{eqnarray}}
\newcommand{\eea}{\end{eqnarray}}

\usepackage{bm}

\begin{document}

\title[Transfer of optical signals around bends in 2D linear photonic networks]{Transfer of optical signals around bends in two-dimensional linear photonic networks}

\author{G M Nikolopoulos}

\address{Institute of Electronic Structure \& Laser, FORTH, P.O.Box 1385, GR-70013 Heraklion, Greece}

\ead{nikolg@iesl.forth.gr}

\begin{abstract}

The ability  to navigate light signals in two-dimensional networks of  waveguide arrays is a prerequisite for  the development of all-optical integrated circuits for information processing and networking. In this article, we present a theoretical analysis of bending losses in linear photonic lattices with engineered couplings, and discuss possible ways for their minimization. In contrast to previous work in the field, the lattices under consideration operate in the linear regime, in the sense that discrete solitons cannot exist. The present results suggest that  the  functionality of linear waveguide  networks can be extended to operations that go beyond the recently demonstrated point-to-point transfer of signals, such as blocking, routing, logic functions, etc.  

\end{abstract}

\maketitle

\section{Introduction}

Photonic lattices (PLs) are currently at the focus of extensive research for two main reasons. Firstly, for their flexibility in simulating various phenomena, especially those related to tight-binding Hamiltonians \cite{PLRev1,PLRev2,PLRev3,PLRev5}, and secondly for their potentials 
as building blocks of quantum circuits for all-optical information processing and networking \cite{3DPLs,PLRev4,SoS}, 
and their role in related studies on quantum random walks \cite{PLrandomWalk} and boson sampling \cite{PLbosonFilter}.  

PLs can be fabricated in a doped silica multilayer structure on a silicon substrate \cite{PLRev4,SoS}, as well as by means of femtosecond laser-writing techniques in the bulk  of glasses \cite{3DPLs,PLRev3}. Both of these techniques allow one to exploit Kerr nonlinearity in order to achieve certain  tasks. In addition, fabrication of waveguides in  LiNbO$_3$ and KTP by means of  etching techniques \cite{PDC} allows for integrated sources of non-classical light, paving thus the way toward integrated quantum chips, where the generation of entangled photons, their transmission, and their processing take place on the same chip. 

The faithful transfer of signals  is a necessary precondition for further developments in these directions, and the problem has attracted considerable interest in recent years \cite{Segment,Lon10,PSTrealization,JogPRA,FSTrealization,protocol,Gor04,prl08,ReviewsPST,EdVolume,HamEng,GenericBends}. 
The typical evolution of an initially well-localized wavepacket  in an ideal finite PL (i.e.,  in the absence of disorder and dissipation), is characterized by spreading,  reflections from the boundaries, and interference phenomena that give rise to diffraction effects \cite{Lon10}. 
At relatively high intensities \cite{remark1}, such distortion effects can be avoided by using discrete solitons as information carriers \cite{PLRev1,PLRev2,PLRev3,3DPLs}. Alternatively, in the linear regime (i.e., for input powers below the threshold for the existence of discrete solitons \cite{remark1}), one may resort to  the segmentation of appropriate lattice sites \cite{Segment}, or to the engineering of judicious couplings between adjacent sites \cite{Lon10,PSTrealization,JogPRA,FSTrealization,protocol,Gor04,prl08}. The latter scenario has been also studied thoroughly in the context of quantum networks, and various solutions have been proposed \cite{ReviewsPST,EdVolume, HamEng, GenericBends}.

In any case, however,  the reliable transfer of signals between two distant nodes of a network 
is not sufficient for large-scale  information processing and networking. To this end, one has to be able to perform reliably and efficiently more complex signal manipulations such as routing, splitting, switching, etc. Such communication tasks can be performed only in higher-dimensional geometric arrangements, where the presence of  bends at different angles are inevitable \cite{3DPLs,Chr01,ExpBend,prl08,GenericBends}.  

The transfer of  signals around bent nonlinear (discrete-soliton) PLs has been investigated to some extent in the literature \cite{3DPLs,ExpBend,Chr01}. On the contrary, the only analogous study for the case of linear networks with engineered couplings has been limited to a generic theoretical model and small bend angles \cite{GenericBends}, so that the effects of bending  can be treated as a small perturbation  to the unbent chain. The regime of bend angles for which this approximation is justified could not be assessed within the generic model of Ref.  \cite{GenericBends}, since it depends strongly on the details of the physical system under consideration. One of the purposes of the present work is to address this question, in the framework of a considerably more elaborate theoretical model that pertains to a two dimensional (2D)  linear PL with engineered couplings, taking into account possible anisotropy effects. To mimic the asymmetry of the waveguides, typically present in experimental realizations, we consider waveguides of an asymmetric rectangular shape.  Moreover, in contrast to Ref.  \cite{GenericBends}, our formalism takes into account all the couplings beyond nearest-neighbours, between any two waveguides. The parameters used throughout our simulations are typical for PLs written in the bulk of glasses,  as this technique has the advantage of being maskless, fast, and rather versatile, allowing for fabrication of optical circuits with 3D layouts, and  waveguides of  controllable transverse profile.  Furthermore,  the  specific coupling configuration we consider, has been implemented recently in this experimental set-up \cite{PSTrealization}. 

The paper is organized as follows.  Our theoretical model, together with various aspects 
of the system under consideration, is presented in Sec. \ref{sec2}. Section \ref{sec3} is devoted to our simulations, with an extensive discussion of our results.   
It is shown that by engineering the couplings between nearest-neighbours one can achieve   faithful transfer of signals between the first and the last  waveguide of a bent PL that operates in the linear regime, for bend angles at least up to $90^\circ$. For sharper bends, the detrimental effects of the bending become very pronounced rapidly, especially in the case of asymmetric waveguides, and can be suppressed by introducing a defect at the corner site, while keeping all the other parameters in the lattice constant. Our simulations suggest that in this way, for the particular coupling configuration under consideration, faithful transfer of signals for bend angles up to $60^\circ$ ($120^\circ$ with respect to the unbent chain), is possible. Qualitative as well as quantitative aspects of the defect required for suppression of  bending effects at different angles are also discussed, and our main results are compared to related results for nonlinear networks, that rely on solitonic information carriers. In the last Sec. \ref{sec4}, we summarize our findings and discuss certain issues that remain open.

\begin{figure}
\centering
\includegraphics[scale=0.9]{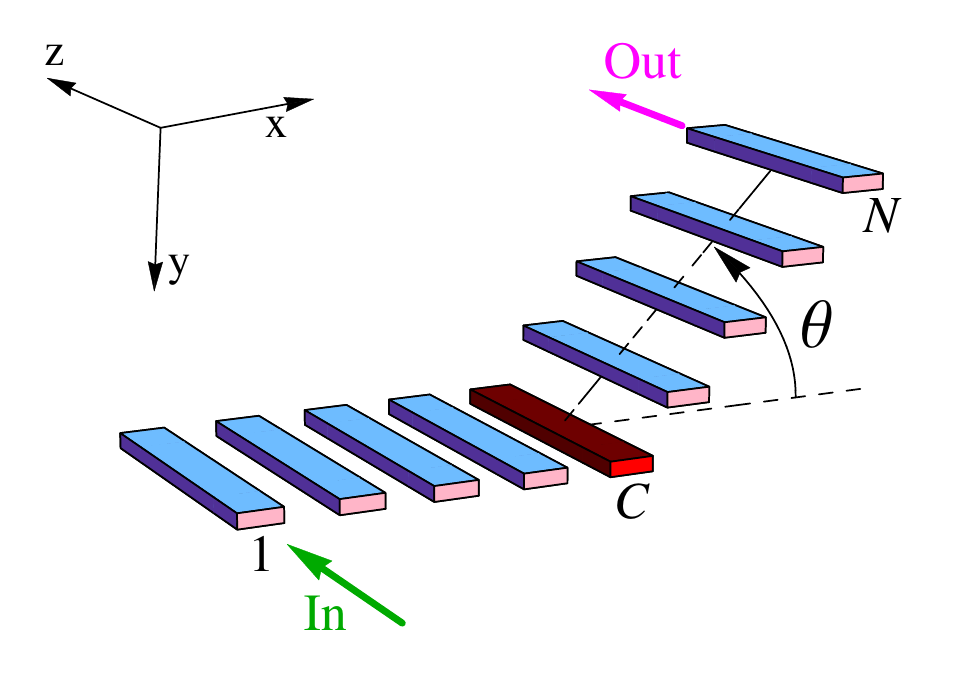}
\includegraphics[scale=0.8]{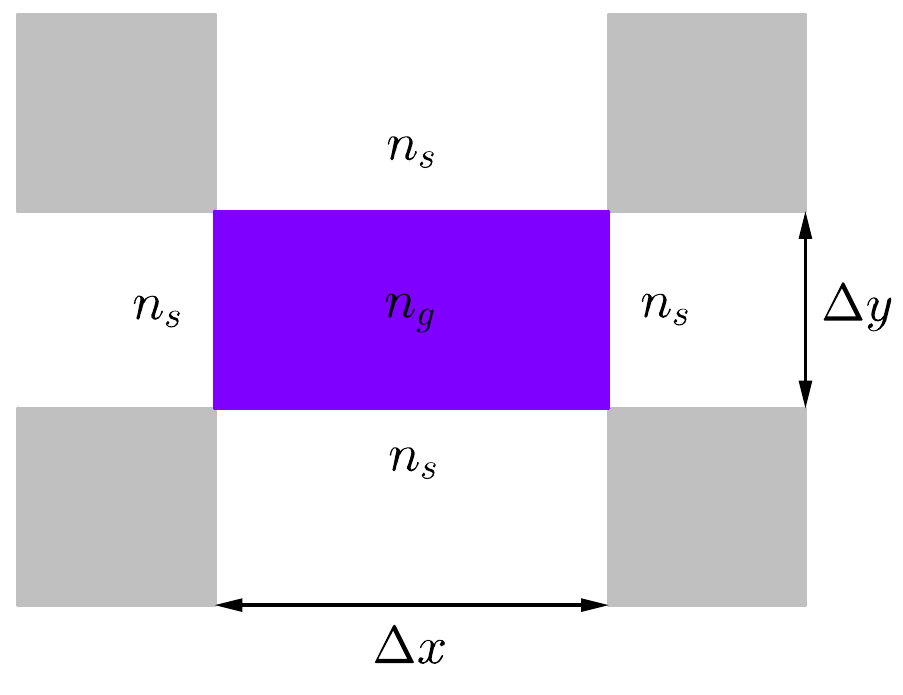}
\caption{(Color online) 
The system under consideration: a bent quantum chain of $N$ rectangular waveguides. Light enters the structure at $z=0$ in the first waveguide, and as it propagates along $z$ it couples to other waveguides.  
Our task is the faithful (ideally perfect) transfer of the input signal to the $N$th waveguide at the exit $(z=L)$ of the structure. The lower scheme shows a cross section of a rectangular waveguide of area $\Delta x\times \Delta y$, 
and refractive index  $n_g$. }
\label{fig1}
\end{figure}

\section{Physical system and modelling}
\label{sec2}
Various 2D configurations of laser-written buried photonic lattices in glasses, have been demonstrated and studied experimentally \cite{PLRev1,PLRev2,PLRev3,3DPLs,ExpBend,PSTrealization}. Typically, the wavenumber along the propagation direction for each waveguide can be  controlled by adjusting the corresponding size and/or the refractive-index change. Moreover, the coupling between two neighboring waveguides drops exponentially with their separation, and the precise form of the exponential law is determined by the details of the experimental set-up. Knowing the precise form of this exponential  law one can engineer various configurations of waveguides that perform certain tasks, such as the non-dispersive transfer of signals between  two waveguides of a PL, for input light with specific properties (i.e., wavelength, polarization) \cite{PSTrealization,FSTrealization}. In most cases, experimental observations have been shown to be in excellent agreement with the predictions of coupled mode theory, and other theoretical models that rely on the Helmhotlz equation for scalar fields.
 
\begin{figure}
\centerline{\includegraphics[scale=0.55]{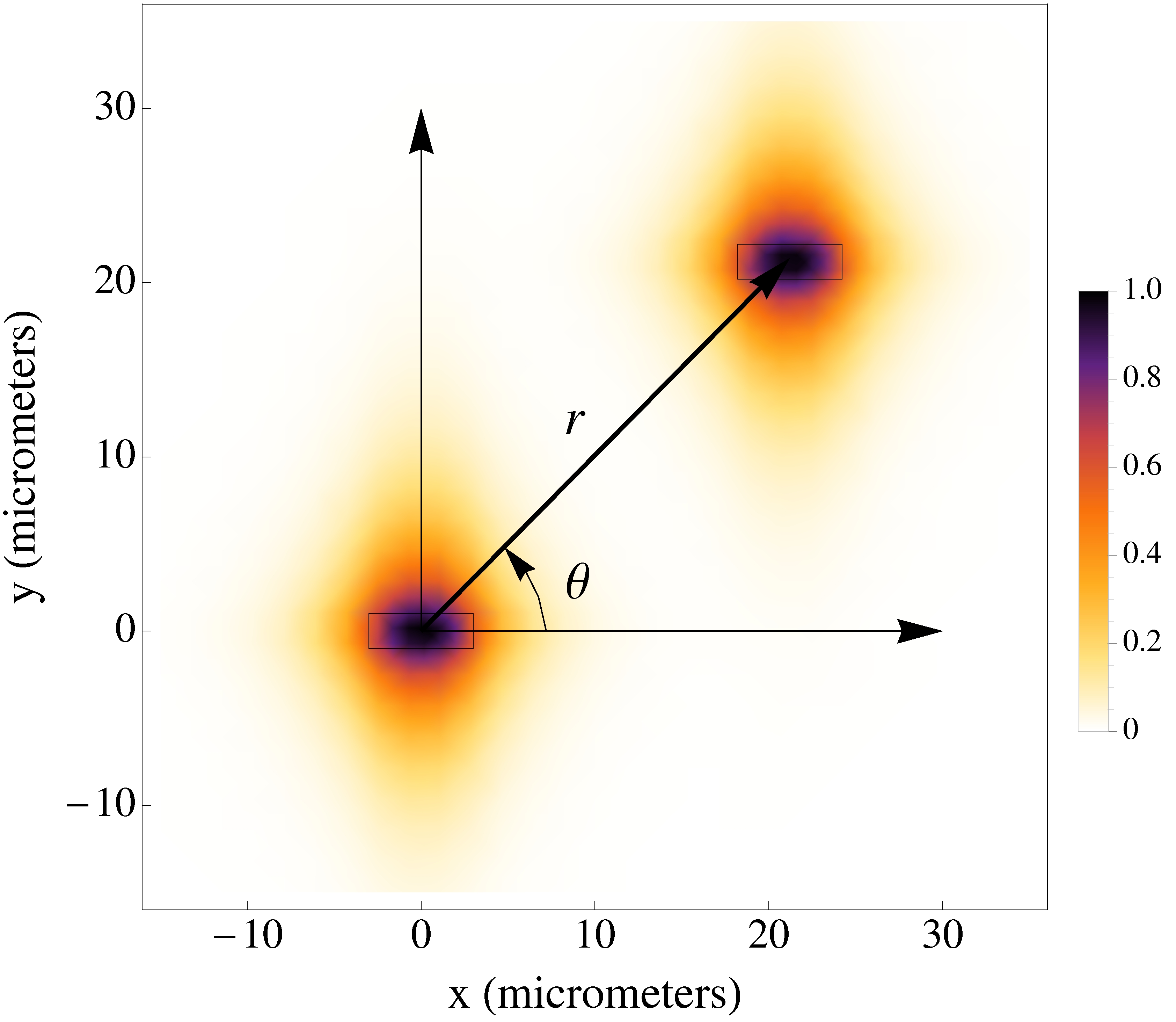}}
\caption{(Color online) 
Eigenmodes of two rectangular waveguides and the geometry of their overlap. 
The two identical waveguides (see rectangles) are shown together with the 
electric field ${\mathcal E}_x^{(j)}(x,y)$. Parameters:   $\Delta x = 6\mu$m, $\Delta y =2\mu$m, $n_s=1.444$, $\delta n = 10^{-3}$, $\lambda = 800 $nm, $r = 30\mu$m, $\theta = \pi/4$.}
\label{fig2}
\end{figure}

In the present work we are interested in the transfer of signals between the two outermost waveguides of a bent array of linear waveguides. 
The 2D arrangement under consideration is depicted in Fig. \ref{fig1}, and pertains to $N$ identical rectangular waveguides, of cross section $\sigma = \Delta x \times \Delta y$ and length $L$.  
Light of specific wavelength $\lambda$, is injected in the first waveguide  at $z=0$, and couples to the neighbouring waveguides as it propagates along $z$. 
Our task is to achieve faithful (ideally perfect) transfer of the signal from the first to the $N$th waveguide at the exit of the 
structure i.e., at $z=L$, for a given value of the permanent bend angle $\theta\geq 0$.  

To this end, for a given value of $\theta$ we engineer the distances between neighbouring  waveguides aiming ideally at a configuration of the coupling constants of the form 
\bea
G_{m,l} = \left \{
\begin{array}{ll} 
\frac{\pi}{2L} \sqrt{(N-M)M}, &   |m-l|=1\\
0, &  \textrm{otherwise}, 
\end{array} \right.
\label{coupling_j}
\eea
for $m,l \in[1,N]$ and $M\equiv\min\{m,l\}$. This is a centrosymmetric configuration with respect to the central waveguide(s) of the PL \cite{ReviewsPST,EdVolume,protocol,Gor04,prl08}. In practise, the 
couplings beyond nearest neighbors are never zero, but they must be negligible relative to the 
nearest-neighbor couplings.  For an unbent PL $(\theta = 0)$, such engineering has been demonstrated recently by two different groups in the framework of PLs in the bulk of fused silica \cite{PSTrealization}. It requires knowledge mainly on the spatial dependence of the coupling, as well as on the refractive index profile for each waveguide  and the form of the excited eigenmodes at a given operation wavelength $\lambda$.  In the following we will investigate whether the coupling configuration (\ref{coupling_j}) can be  implemented in the same manner,  in bent PLs (i.e., for $\theta >0$), with symmetric or asymmetric waveguides. Before we address this question, let us present our model for the waveguides.  

\subsection{Eigenmodes of independent waveguides}
In various physical realizations the waveguides have an asymmetric elliptic profile, which is reflected in the observed modal field distribution at a given wavelength and field polarization, as well as in the dependence of the coupling on the waveguide separation \cite{PLRev3,3DPLs,FSTrealization,AsymWG}. Although, as discussed in Sec. \ref{sec4}, the design of symmetric waveguides is possible with current technology, throughout the present work we present results for waveguides with  asymmetric shape, since in this case the presence of bends turns out to be more pronounced than for symmetric waveguides.  Analogous simulations for symmetric waveguides have also been performed, and we will refer to related findings wherever necessary, pointing out the main differences from the case of asymmetric waveguides.  

We consider  rectangular waveguides with $\Delta x \geq \Delta y$, corresponding to the major and minor axes of the  elliptic profiles typically observed experimentally.  For the analysis of the waveguides we follow the Marcatili's approach, which is widely used in photonics and optoelectronics \cite{Marc69,book1,book2}. A cross-sectional view of the waveguides under consideration is shown in the lower Fig. \ref{fig1}, where $n_g$ and $n_s$ denote the refractive indices for the core and the cladding (substrate) respectively. Typically, their difference is very small $(\sim 10^{-3})$ and thus  the refractive index distribution for the $j$th rectangular  waveguide is well approximated by 
\bea
n_j^2(x,y) \approx n_{j;x}^2(x)+n_{j;y}^2(y) + {\cal O}(n_g^2 - n_s^2)
\eea
with 
\bea
 n_{j;x}^2(x) = \left \{
\begin{array}{ll} 
n_g^2/2, &  |x| \leq \Delta x/2\\
n_s^2-n_g^2/2, &  |x| > \Delta x/2 
\end{array} \right.
\eea
and 
\bea
 n_{j;y}^2(y) = \left \{
\begin{array}{ll} 
n_g^2/2, &  |y| \leq \Delta y/2\\
n_s^2-n_g^2/2, &  |y| > \Delta y/2 
\end{array} \right. ,
\eea
where we have assumed that the waveguide is centred at $(x,y)=(0,0)$.
In the following, for the sake of simplicity and without loss of generality we set 
\bea
n_g = \frac{n_s}{\sqrt{1-2\delta n}}
\eea
where $\delta n$ is the modification of the refractive index. 
The refractive index of the shaded (corner) areas in the lower scheme of Fig. \ref{fig1}, is approximated by $\sqrt{2n_s^2-n_g^2}\approx n_s(1-\delta n-{\cal O}(\delta n^2))$. Clearly, for $\delta n\sim 10^3$, one has $\sqrt{2n_s^2-n_g^2}\approx n_s$. 

The dimensions of the waveguides and the associated  refractive-index modulations adopted throughout this work are within the range of values one typically  
finds in experiments pertaining to laser-written waveguides in  glasses.  There are some quantitative differences, however, since the adopted rectangular profile is not expected to capture precisely all the features of the modes observed in experiments (e.g., precise form of modes, 
 penetration depth, etc). One of the key features in most of the experiments is that only one (the lowest) eigenmode is excited at the operation wavelength. Hence, the waveguide parameters we consider here are such that the lowest mode of the rectangular waveguide is  excited  (i.e., $\mathbb{E}_{1,1}^{(x)}$	or $\mathbb{E}_{1,1}^{(y)}$), which means that the electric field exhibits only one  peak along both x- and y-axis directions. The main field components for modes $\mathbb{E}_{1,1}^{[x(y)]}$ in the $j$th waveguide are  $E_{x(y)}^{(j)}$  and $H_{y(x)}^{(j)}$,  with  the electric field polarized along the $x(y)$ direction, respectively.
 
The focus of the present work is on the effects of bends and to this end 
we ignore various types of possible imperfections so that all the estimated losses in our simulations can be attributed only to bending effects. Furthermore it is sufficient to consider the  mode $\mathbb{E}_{1,1}^{(x)} $ in  the following analysis, since the calculations for the $\mathbb{E}_{1,1}^{(y)}$  mode are the same. 
We will return to this point in the concluding remarks of the present work. 

The components of the electric $({\bm E}^{(j)})$ and magnetic $({\bm H}^{(j)})$ fields that prevail in the $j$th waveguide are \cite{Marc69,book1,book2}
\bea
&& E_x^{(j)} = {\mathscr E}_{x}^{(j)}(x,y) \exp[i(\omega t+\beta_j z)] \\
&& H_y^{(j)} = {\mathscr H}_{y}^{(j)}(x,y) \exp[i(\omega t+\beta_j z)],
\eea
 whereas $H_x^{(j)} = 0$. From Maxwell's equations one has: 
\bea
\label{Hy_eq}
&&\frac{\partial^2 {\mathscr H}_y^{(j)}}{\partial x^2}+ \frac{\partial^2 {\mathscr H}_y^{(j)}}{\partial y^2} 
+ [k^2 n_j(x,y)^2 - \beta_j^2] {\mathscr H}_y^{(j)} = 0,\\
&& {\mathscr E}_x^{(j)} \approx \frac{\omega\mu_0}{\beta_j} {\mathscr H}_y^{(j)},
\label{Ex_eq}
\eea
where $\omega$ and $k$ are the frequency and the wavenumber of the input light, whereas $\beta_j$ is the 
wavenumber along the propagation direction of the waveguide (z-axis). 
Equation (\ref{Hy_eq}) can be solved numerically, or analytically for the model under consideration (see appendix).  In Fig. \ref{fig2} we show 
the electric field  ${\mathscr E}_x(x,y)$ as obtained in our simulations, for a particular set of parameters. Clearly, the asymmetry of the waveguides is also reflected in the eigenmodes, and as will be seen later on, it also affects the coupling between adjacent waveguides. It has to be emphasized here, however, that the depicted modal profile is for the sake of illustration only, and pertains to the particular parameters given in the caption. The quantitative aspects of the 
modal profile may change e.g., by changing the dimensions of the waveguide, the wavelength of the light, etc.  

Finally, the normalization we have adopted  throughout this work implies that the power carried by the eigenmode  of the $j$th waveguide along the propagation direction is \cite{EdVolume,book1,book2, book3} 
\bea
P_j = \frac{1}{2}\int \int \Re \left [({\bm E}^{(j)}\times {\bm H}^{(j)\star} )\cdot \hat{\bm z}\right ]\,  dx dy = 1\,\textrm{(Watt)}. 
\eea  

\subsection{Coupled-mode theory}

As shown in Fig. \ref{fig2}, when two waveguides are brought close together, their  optical modes overlap. For sufficiently small overlaps, the electromagnetic field distribution for either of the neighbouring waveguide does not differ substantially from the one for an isolated waveguide, and the propagation characteristics of the coupled waveguides can be analyzed by means of  the coupled-mode theory, the details of which can be found almost in every textbook on photonics and optoelectronics (e.g., see \cite{EdVolume,book1,book2, book3}). 
For the sake of completeness, here we sketch the main steps of the approach. 

The total electric field in a configuration of $N$ evanescently coupled identical waveguides  is well approximated by the superposition 
\bea
{\bm E}(x,y,z) = \sum_{j=1}^N A_j(z) {\mathscr E}_x^{(j)}(x,y)\exp[i(\omega t - \beta_j z)] \hat{\bm x},
\label{totalEF}
\eea
where ${\mathcal E}_x^{(j)}$ are determined by Eq. (\ref{Ex_eq}).
Substituting  this expression into the wave-equation 
\bea
\left [ {\nabla}^2
+ \frac{\omega^2}{c^2}n^2(x,y)
\right ] {\bm E}(x,y,z)  = 0, 
\eea
where $n(x,y)$ the refractive-index distribution of the entire  structure  of the coupled 
waveguides, and following standard well-known steps one obtains a closed set of differential equations for the amplitudes $A_j(z)$
\bea
\frac{d{\mathfrak A}}{dz} = {\mathfrak  J}\cdot  {\mathfrak A}
\label{eom}
\eea
where ${\mathfrak A}\equiv (A_1, A_2, \ldots, A_N)^T$ and ${\mathfrak  J}$ is an $N\times N$  matrix with all the diagonal elements equal to zero, and the off-diagonal elements given by 
\bea
{\mathfrak  J}_{m,l} = {\mathscr  J}_{m,l}\exp[i(\beta_m-\beta_l)z],
\eea
with  the coupling between the $m$th and the $l$th waveguide  given by 
\bea
 {\mathscr  J}_{m,l} = \frac{\omega \varepsilon_0}{4} \int \int dx dy\,
{\mathscr E}_x^{(m)*}(x,y)  \Delta n_l^2 (x,y) {\mathscr E}_x^{(l)}(x,y),\nonumber\\
\label{eq:coupling}
\eea
where $\Delta n_l^2\equiv n^2(x,y) - n_l^2(x,y)$, with $n_l(x,y)$ the refractive-index  profile for the $l$th waveguide alone. 

Before we focus on the behaviour of the couplings for the particular setup under consideration,  it is worth mentioning that in addition to the coupling between adjacent waveguides, in the framework of coupled-mode theory one also obtains corrections to the propagation wavevector $\beta_l$ of the $l$th waveguide, due to the presence of the adjacent waveguides, as well as the so-called 
``butt-coupling" coefficients. Such terms are typically much smaller than ${\mathscr  J}_{m,l}$ and thus their effects are neglected here \cite{book1}.  

\subsection{Coupling constants}
\begin{figure}
\centerline{\includegraphics[scale=0.37]{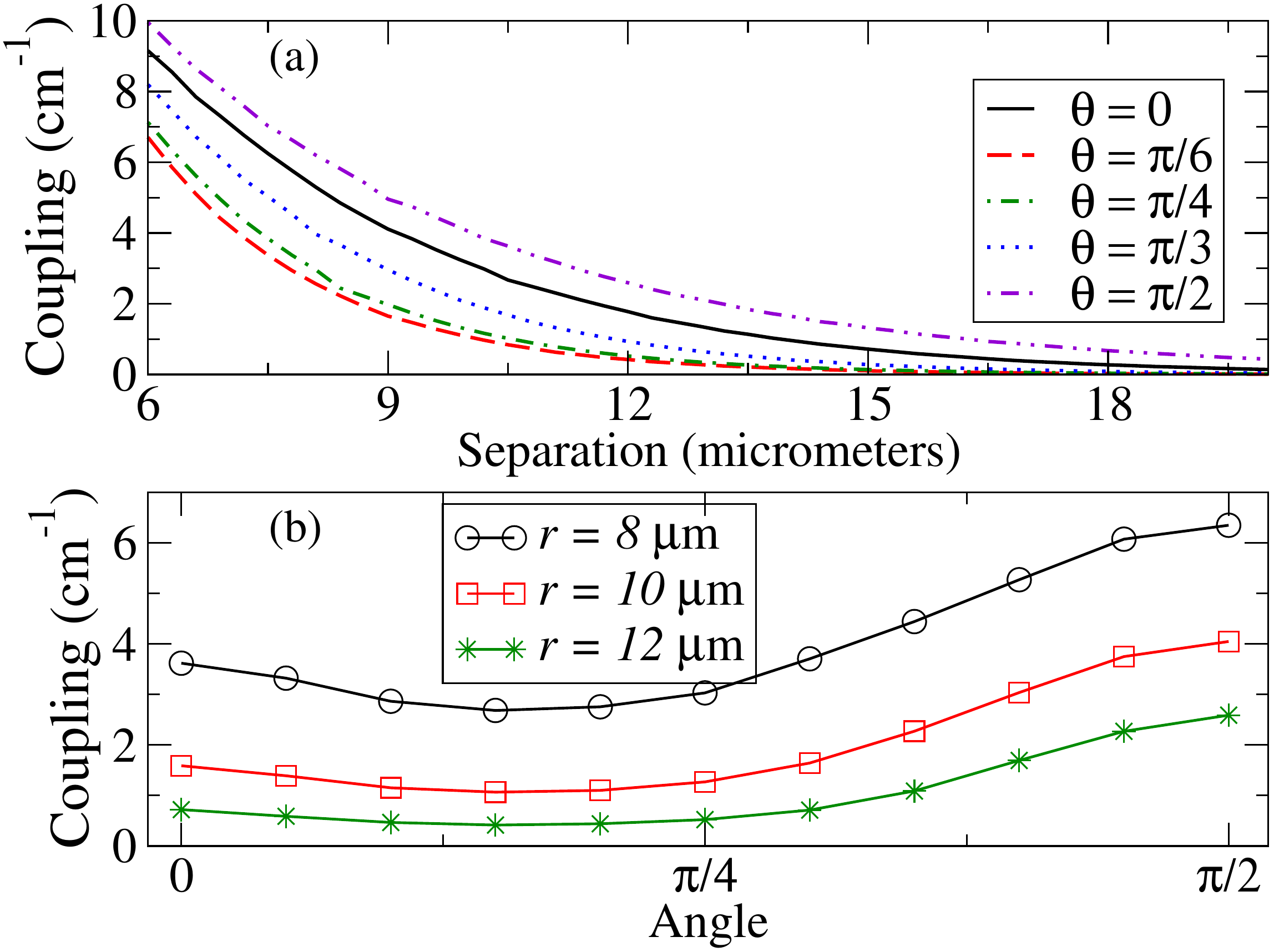}}
\caption{(Color online) 
Geometry of the coupling between two rectangular waveguides (see Fig. \ref{fig2}). 
(a) The coupling as a function of the separation $r$ at different angles $\theta$. 
(b) The coupling as a function of the angle $\theta$ at different separations $r$. 
Other parameters as in Fig. \ref{fig2}. 
}
\label{fig3}
\end{figure}

Equation (\ref{eq:coupling}), shows that the coupling between two waveguides originates from the overlap between the corresponding 
eigenmodes. Hence, one expects the asymmetry of the eigenmodes  to be also reflected in the couplings. Consider 
the directional coupler of Fig. \ref{fig2}. The corresponding coupling between the two waveguides is plotted  in Fig. \ref{fig3}(a) as a function  of the separation $r$ 
(measured for the centers of the waveguides), at a fixed angle $\theta$. For any value of $\theta$ the  dependence of the coupling on $r$ is well approximated by an exponential of the form 
\bea
 {\mathscr  J}_{m,l} (r,\theta) = \mu(\theta)\exp[-\xi(\theta) r],  
 \label{exp_law}
\eea 
and it is anisotropic since the details of the exponential drop with increasing $r$ depend on $\theta$. Indeed, as shown in Fig. \ref{fig3}(b), for fixed $r$ the coupling varies by a factor of 3, as we change the angle $\theta$ from $0$ to $\pi/2$. The relative position of the curves for different $\theta$ in Fig. \ref{fig3}(a), depends strongly on the specific parameters under consideration (e.g., wavelength of light, refractive-index modulation, height and width of the rectangular shape). 
In the case of symmetric waveguides, for example, the spread of the curves for various $\theta$ is considerably smaller \cite{remark3}. Varying the refractive-index profile of the waveguides, one can change the precise form of the modal distribution [see Fig.  \ref{fig2}], and thus the values of the parameters $\mu,\xi$ in Eq. (\ref{exp_law}) for a given $\theta$. 

In any case, the crucial point is that the present model captures the dependence of the coupling  on both $r$ and $\theta$ (see Eq. \ref{exp_law}), which is also what one has in practise. The precise mathematical form of this dependence on $\theta$ is not crucial for what follows and actually, it is never used explicitly. The key point is that such a type of anisotropy allows for the engineering of couplings at different coupling angles $\theta$, by adjusting the distance $r$ (e.g., see \cite{AsymWG}).  For the implementation of a centrosymmetric  coupling configuration [such as the one in Eq. (\ref{coupling_j})] in a bent chain with $\theta > 0$, the separations $r_{j,j+1}$ between waveguides with indices below the index of the corner site $C$, have to be different from the separations for  waveguides with indices above $C$. In other words, a centrosymmetric coupling configuration does not imply centrosymmetric distribution of the separations $r_{j, j+1}$  in the case of an anisotropic spatial dependence of the coupling.

\begin{figure*}
\centering
\includegraphics[scale=0.8]{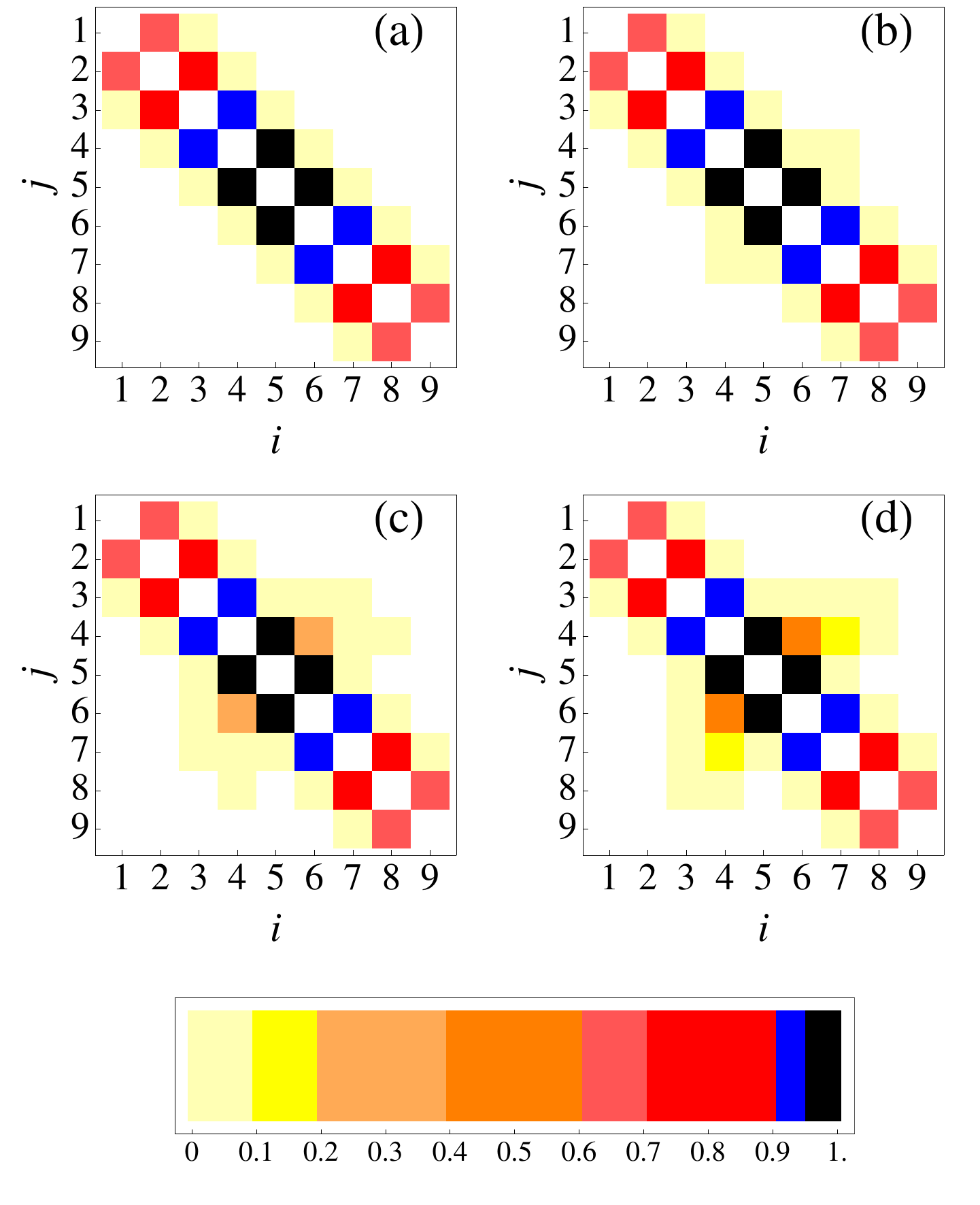}
\caption{(Color online) The relative strength of the couplings between different identical waveguides ${\mathscr J}_{i,j} [\max{\{{\mathscr J}_{i,j}\}}]^{-1}$,  for a bent chain of $N=9$ waveguides with $r_{j,j\pm 1}$ such that ${\mathscr J}_{j,j\pm 1}$is given by Eq. (\ref{coupling_j}), 
and bend angles: (a) $\theta = 0$, (b) $\theta = 16 \pi/32$, 
(c) $\theta = 19 \pi/32$, (d) $\theta = 20 \pi/32$. Other parameters as in Fig. \ref{fig2}. }
\label{fig4}
\end{figure*}

\begin{figure*}
\centering
\includegraphics[scale=1,clip]{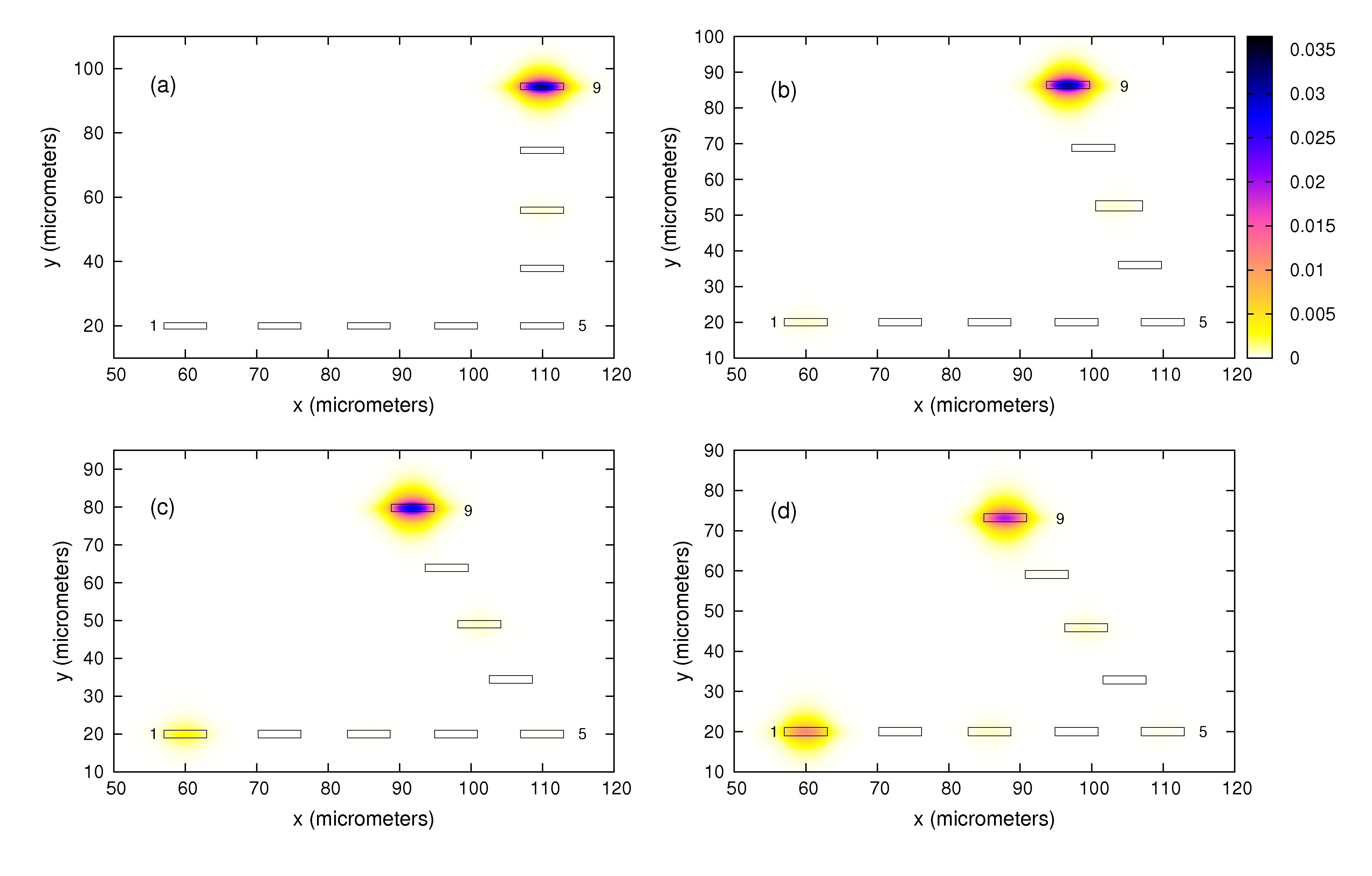}
\caption{(Color online) Intensity distribution at the output ($L=10$ cm) of a bent chain with $N=9$ identical 
rectangular waveguides (also shown on each plot).  Bend angles: (a) $\theta = 16 \pi/32$,  (b) $\theta = 18 \pi/32$, 
(c) $\theta = 19 \pi/32$, (d) $\theta = 20 \pi/32$. The intensity is measured in units Wcm$^{-2}$ and the total power 
in the sample at any $z$ is normalized to the input power.  Other parameters as in Fig. \ref{fig2}. Note the 
different scale in the $y$ axis.}
\label{fig5}
\end{figure*}

\section{Light transport through bent photonic lattices}
\label{sec3}

Throughout our simulations we worked on a three-dimensional grid in a sufficiently large box so that reflections from the boundaries are absent. The total electric field at a given point  $(x,y,z)$ 
was estimated according to Eq. (\ref{totalEF}), through the  solution of Eqs. (\ref{Hy_eq}), (\ref{Ex_eq})  and (\ref{eom}).  Working with different parameters,  we have reached similar conclusions and for the sake of concreteness, in this section we present results  pertaining to a bent PL consisting of $N=9$ nearly identical waveguides of length $L=10$ cm. 
The waveguides are written in the bulk of a glass with $n_s=1.444$ and the associated  refractive index change is $\delta n = 10^{-3}$, whereas their cross-section is $\sigma=6\times 2\mu{\rm m}^{2}$.  For a given $\theta\geq 0$, the distances between successive waveguides $r_{j,j\pm 1}$ are engineered  so that  the nearest-neighbour couplings ${\mathscr J}_{j,j\pm 1}$ are well approximated by Eq.  (\ref{coupling_j}) for $L=10$ cm and $N=9$.  Light of wavelength $\lambda = 800$ nm, and sufficiently low power so that nonlinear effects are negligible,  is injected in the first waveguide at $z=0$, and couples to the other waveguides as it propagates along $z$. 

\subsection{Coupling constants}

From a theoretical point of view, the propagation of light in the PL is determined by Eqs. (\ref{eom}), and in particular by the coupling matrix ${\mathfrak J}$.  When the distance between non-neighbouring waveguides is sufficiently large, couplings beyond nearest neighbours  are negligible and thus the coupling matrix ${\mathfrak J}$  has basically tridiagonal form. Recent experiments on the realization of the coupling configuration (\ref{coupling_j}) in unbent PLs  $(\theta = 0)$ have shown that the assumptions underlying the coupled-mode theory, as well as the assumption of nearest-neighbour couplings of the form (\ref{coupling_j}) can be fulfilled experimentally for a moderate number of waveguides, and thus faithful transfer between the two ends of the unbent chain has been observed \cite{PSTrealization}. For an unbent chain of a given length, these assumptions are expected to break down only for a large number of waveguides  (e.g., see related discussion in the work of Bellec {\em et al}. \cite{PSTrealization}). 

By contrast, the realization of the coupling configuration (\ref{coupling_j}) for bent chains  with $\theta > 0$ has not been investigated in the literature so far. In particular, given that 
non-neighbouring waveguides around the corner $(C)$ come closer as one increases $\theta$ (see Fig. \ref{fig1}), 
couplings beyond nearest neighbours are expected to increase. There should 
exist, therefore, a critical angle $\theta_c$ above which the couplings beyond nearest neighbours become comparable to the nearest-neighbour ones, and thus their effects cannot be neglected.  Our first task here is to estimate the critical angle for the 
particular set-up under consideration.  Subsequently, for  angles $\theta>\theta_c$ our task is 
to investigate whether it is possible to improve the transfer of the signal between the two outermost waveguides of the PL, without additional extensive engineering \cite{remark4}.

For the reasons explained above, in our formalism the matrix ${\mathfrak J}$ includes the couplings for all possible pairs of waveguides. The relative strengths of the couplings in the matrix ${\mathfrak J}$  (with $\beta_m=\beta_l\,\forall m,l$) are plotted  in Fig. \ref{fig4} for a chain of $N=9$ waveguides, and  for increasing bend angles, with the corner site $C=5$. For the unbent chain $(\theta = 0)$ as well as for $\theta < \pi/2$, the couplings beyond nearest neighbours are at 
least one order of magnitude smaller than the nearest-neighbour couplings, and thus they can be safely ignored. For  $\theta = \pi/2$ the strength of the couplings between the waveguides $C- 1$ and $C+1$  has been doubled, whereas couplings 
between higher-order neighbours emerge. For $\theta = 19\pi/32$ and $\theta = 20\pi/32$  the couplings between the waveguides $C- 1$ and $C+1$ are comparable to the nearest-neighbour couplings, and the couplings between higher-order neighbours also increase.  These observations suggest that strong deviations from the coupling configuration of the unbent chain are expected for bend angles above $\theta_c=90^\circ$, whereas the deviations for  angles up to $90^\circ$ are not expected to be so pronounced \cite{remark5}. 

In the case of symmetric waveguides (i.e., for $\Delta x = \Delta y = 6\mu$m) our simulations show that the couplings beyond nearest neighbours are less sensitive to bends. For example, for $\theta = 20 \pi/32$ we find that the coupling ${\mathscr J}_{C- 1, C+ 1}$  is at least five times smaller than ${\mathscr J}_{C\pm 1,C}$, whereas couplings between higher-order neighbours can be safely ignored.
This is because the confinement of the lowest eigenmodes in all directions turns out to be  stronger than in the case of the asymmetric waveguides with $\Delta x = 6 \mu$m and $\Delta y = 2\mu$m.

\subsection{Output intensity distributions and losses}

\begin{figure}
\centering
\includegraphics[width=8.5cm,clip]{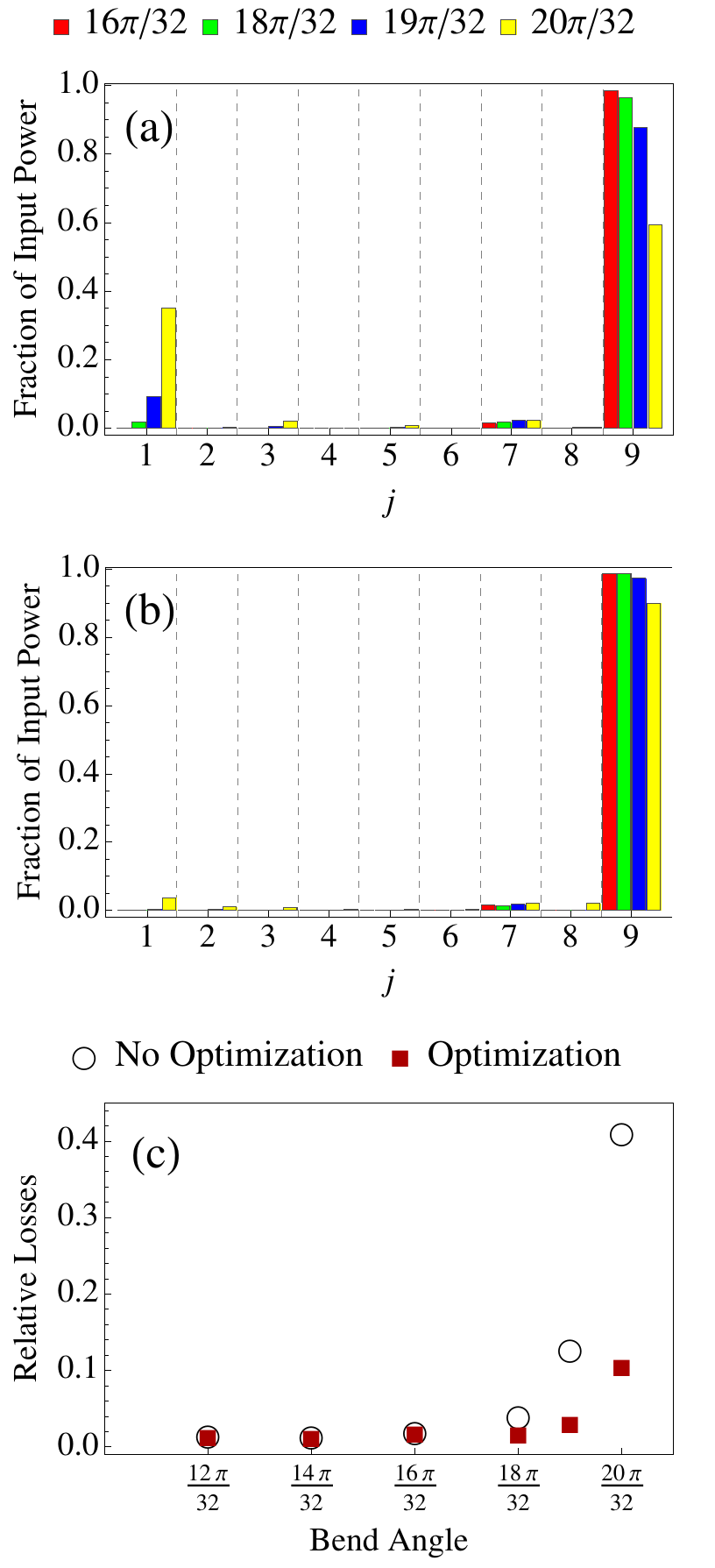}
\caption{(Color online) Losses in the transfer of signals  along a bent chain of $N=9$ identical rectangular waveguides. 
(a) Fraction of the input power that has been transferred to the $j$th waveguide at the output of the sample 
(i.e., at $L=10$ cm ),  for different bend angles. (b) As in (a) with optimized corner site.  (c) The relative losses at the exit of the sample,  and for different bend angles with and without optimized corner. Other parameters as in 
Fig. \ref{fig5}.}
\label{fig6}
\end{figure}

As mentioned before,  one expects ideally complete transfer of the light between the two outermost waveguides, when the matrix elements ${\mathfrak J}_{m,l}$ are well approximated by 
Eq. (\ref{coupling_j}). Our simulations show that this happens for $\theta\leq \pi/2$ and in Fig. \ref{fig5}(a) we show only the intensity distribution at the output  for the case of $\theta=\pi/2$. As we increase $\theta$ further, couplings beyond nearest neighbours distort the transfer between the two outermost waveguides. For the sake of illustration, in Figs. \ref{fig5}(b-d) we present the intensity distributions at the output of a bent chain for  $\theta = 18\pi/32,\,19\pi/32,$ and $20\pi/32$,  respectively.   One can see that the output intensity is not restricted only to the 9th waveguide, but there are also  non-negligible  fractions in other waveguides, including  the 1st and the 7th one. A clearer quantitative picture can be 
obtained by looking at  how the input power is distributed among the waveguides at the output. As shown in Fig. \ref{fig6}(a), 
for $\theta = 19\pi/32,$ and $20\pi/32$ we find that 85\% and 60\% of the input power respectively, has been  transferred to the target 
waveguide at the output, whereas a significant fraction of the  input light  can be found at the exit of all the other waveguides, but mainly of the first one.  Hence, as shown in Fig. \ref{fig6}(c) (see open circles), the relative losses are about 
15\% and 40\%, respectively. 

The present scheme that relies on engineered couplings in linear PLs seems to be a bit more robust against bending losses, than schemes that rely on solitonic signals and nonlinear PLs. For instance, the authors of Ref. \cite{Chr01} have estimated that for $\varphi = \pi-\theta = 90^\circ$  solitons suffer about 
$5\%$ bending losses, whereas for $\varphi = 70^\circ$ the losses exceed  $38\%$.  
As shown in Fig. \ref{fig6}(c), in the present scheme bending losses do not exceed $5\%$ for angles $\theta\lesssim 100^\circ$, whereas for  $\theta \approx 113^\circ$ (corresponding to $\varphi \approx 67^\circ$) bending losses are about $40\%$. For the reasons discussed above, the losses for the same set-up can be reduced considerably by using symmetric waveguides. In this case we find that losses do not exceed 10\% for $\theta \approx 113^\circ$, which suggests that linear PLs with symmetric waveguides considerably outperform nonlinear PLs with solitonic informations carriers.

We turn now to discuss a method for minimizing the bending losses, by introducing a 
defect at the corner site. 

\begin{figure}
\centering
\includegraphics[width=8.5cm,clip]{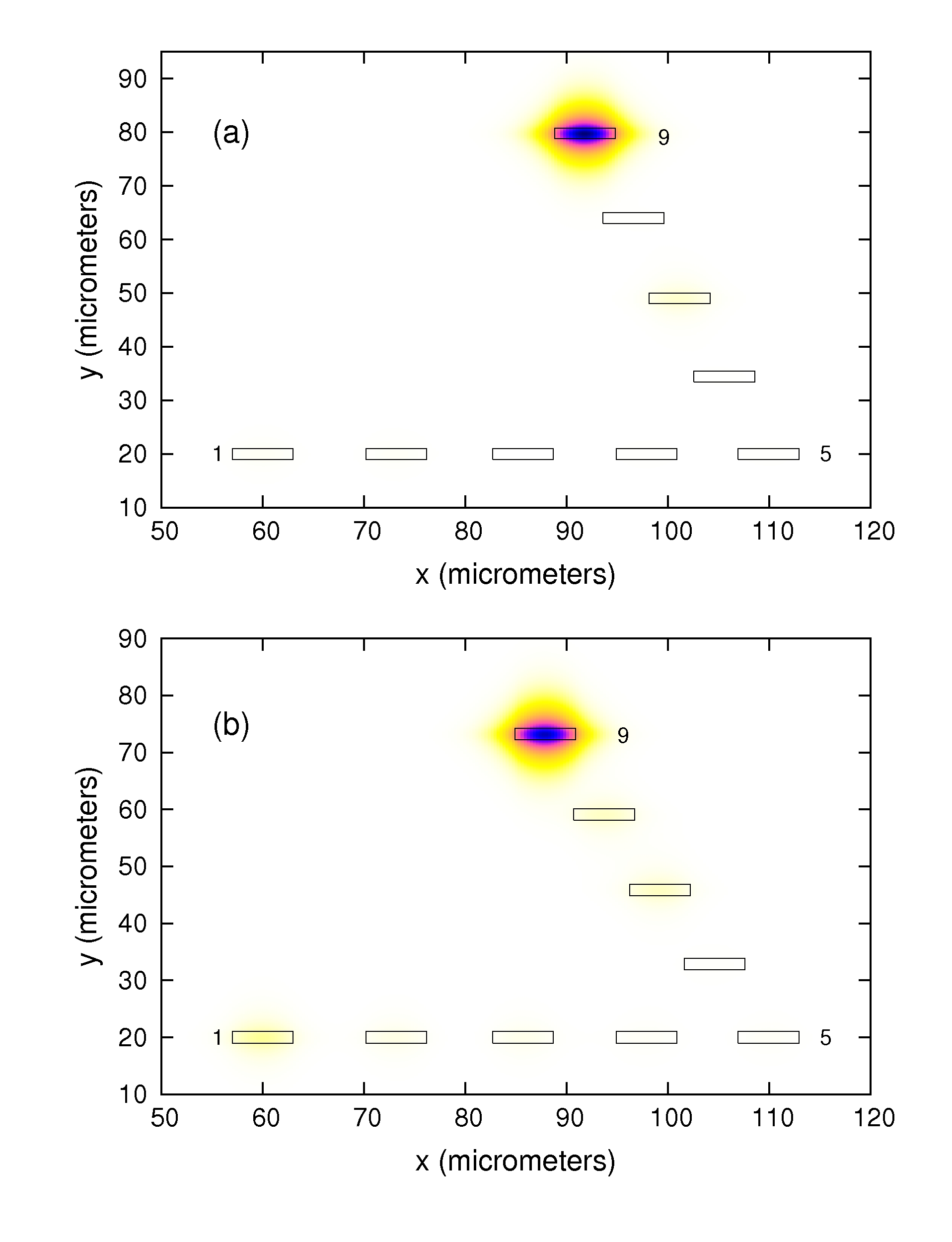}
\caption{(Color online) As in Fig. \ref{fig5} after optimization of the corner site $C=5$, for (a) $\theta = 19 \pi/32$ and (b) $\theta = 20 \pi/32$.  }
\label{fig7}
\end{figure}
 
\subsection{Minimization of bending losses}

In the framework of solitonic signals it has been shown that bending losses can be reduced by  introducing a defect at the corner site of the bend \cite{Chr01,ExpBend}. More recently, it was shown that the same method also works in the framework of linear Hamiltonians with engineered couplings \cite{GenericBends}. The generic model of Ref. \cite{GenericBends}, however,  did not allow for an in depth investigation of various quantitative aspects of the defect. The present model allows us to gain further insight into the method, and shed light on crucial questions pertaining to the size and the  refractive-index change of the defect.  

We assume that all of the waveguides, but the corner one, are identical, and let us denote by $\beta$ the corresponding propagation wavenumber.  The wavenumber for the corner site will be denoted by $\beta_C$ and let  $\Delta\equiv \beta_C-\beta$ be the detuning of the corner site relative to the other waveguides of the chain. 
In our simulations, for a given angle this detuning has been optimised, while keeping all other parameters of the PL fixed, so that the transfer from the first to the last waveguide is maximized (i.e. losses are minimized). 

Figures \ref{fig7}(a,b) show the intensity distribution at the output of a bent PL for two different angles after optimization 
of the wavenumber for the corner waveguide. Comparing these two figures to the corresponding figures 
without optimization [see Fig. \ref{fig5}(c,d)], one sees a clear improvement of the transfer from the first 
to the last waveguide. Still, there are fractions of the input signal that are not found at the exit of  the target waveguide, but certainly they are considerably smaller than in  Figs. \ref{fig5}(c,d). Indeed, as shown in Fig. \ref{fig6}(b), for bend angles up to $20\pi/32$, more than about 90\%  of the input power has been transferred to the target waveguide at the output, whereas the relative losses are at least twice smaller than without optimization and they hardly exceed 10\% [see filled squares in Fig. \ref{fig6}(c)]. We see therefore that by introducing a corner defect one can minimize bending losses for fixed $\theta$ in the present linear array with engineered couplings, but the same approach seems to work more efficiently for nonlinear arrays and solitons. According to Ref. \cite{Chr01} bending losses 
after the inclusion of defect are restricted to less than approximately $1\%$ for 
angles $\theta = 90^\circ$ and $110^\circ$, whereas in our case we find losses approximately 1.4\% for $\theta = 90^\circ$  and $ 5\%$ for $\theta = 110^\circ$.  However, when symmetric waveguides are considered in our scheme, the corresponding bending losses  after optimization do not exceed 1\% for angles up to $\theta = 113^\circ$; a performance that is comparable to (if not better than) the performance of the bent nonlinear PLs considered in Ref. \cite{Chr01}.

The quantitative aspects of the defect are intimately connected to the details of the set-up 
under consideration. The optimal values of the detunings that minimize bending losses at various $\theta$ in our model  are given in table \ref{tab1}. Clearly, in all cases $\Delta$ is negative and increases (in absolute value) as we increase the bend angle.  Typically, the wavenumber of a waveguide can be controlled by changing the size of the waveguide, 
or by adjusting the associated refractive-index modification. An estimation of the changes required to 
achieve some of the detunings discussed here are shown in the last columns of table \ref{tab1}. A close inspection of these values shows that in order to achieve the estimated 
optimal detunings the accuracy required in the writing of the waveguides is at least $10^{-5}$ 
in refractive-index changes and at least $10^{-2}$ cm$^{2}$ in the cross-section of the waveguides. Interestingly enough, the present estimations for the required refractive-index changes are comparable to related estimations for solitonic schemes \cite{Chr01,ExpBend}.

\begin{table}
\caption{\label{tab1} Optimal detunings of the corner site that minimize bending losses at different angles in a linear chain with $N=9$ asymmetric waveguides of length $L=10$ cm. The 
detuning is defined as $\Delta = \beta_C -\beta$ where for the parameters under consideration, $\beta\simeq 11.3392\times10^{4}$ cm$^{-1}$.  The strength of the detuning 
relative to the coupling  $G_{C,C\pm1}$ (see Eq. (\ref{coupling_j})) is given in the third column. The fourth and the fifth column give the control required on the refractive-index modulation and the size of the corner site relative to $\delta n =10^{-3}$ and $\sigma = 12\mu{\rm m}^2$ respectively, in order to achieve the optimal detunings of the second column.}
\footnotesize\rm
\begin{tabular*}{\textwidth}{@{}l*{15} {@{\extracolsep{0pt plus12pt} } c} }
\br
Angle $(\times \pi/32)$ & $\Delta$ (cm$^{-1}$)  &  $|\Delta|/G_{C,C\pm1}$  & Relative change of refractive index$^1$ & Relative change of size$^2$\\
\mr
18 &  -0.1955 &  0.275  & 0.46 & 0.25\\ 
19 &  -0.4317  &  0.608  & 1.60 & 0.85\\ 
20 & -1.0733  &  1.512  & 5.15 & 2.50\\ 
\br
\end{tabular*}
$^1$Defined as $100\times (\delta n-\delta n_C)/\delta n$.\\
$^2$Defined as  $100\times (\sigma-\sigma_C)/\sigma $.
\end{table}

\begin{figure}
\centering
\includegraphics[scale=0.32]{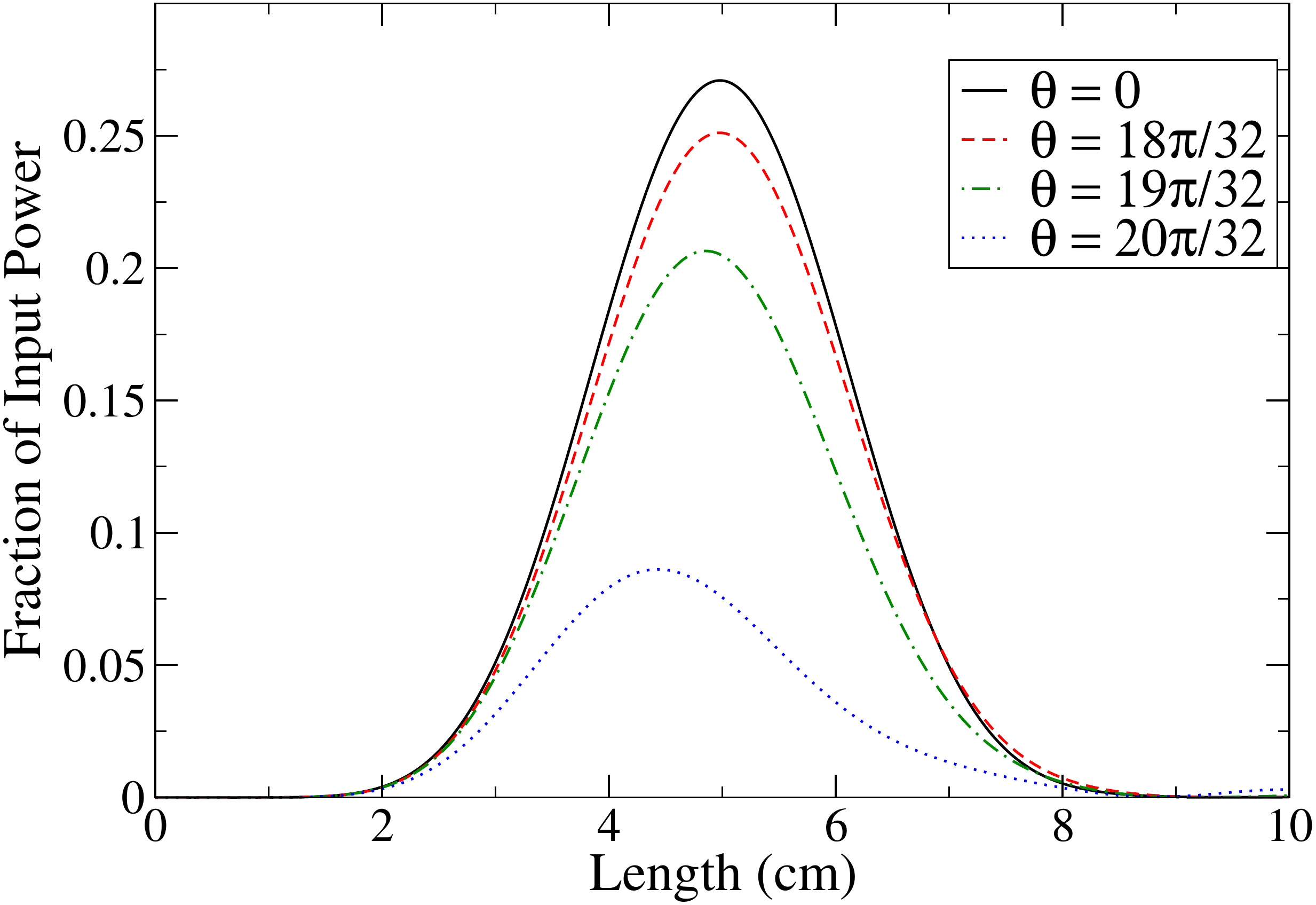}
\caption{(Color online) Fraction of the input power that is found at the corner waveguide  
at different lengths. The detuning of the corner waveguide has been optimized to minimize bending losses. Other parameters as in Fig. \ref{fig5}.
}
\label{fig8}
\end{figure}

Before closing this section it is worth discussing briefly the reason for the success of the corner defect  in minimizing bending losses. 
In the case of solitonic signals and nonlinear PLs it has been conjectured that the detuning 
of the corner site relative to the rest of the lattice virtually  removes the corner 
site from the lattice \cite{ExpBend}. 
Thus, the remaining (identical) waveguides effectively constitute a  
smoother link, which is reflected in the improvement of the transport. This explanation does not apply to our setup as we work in the linear regime and the coupling configuration under consideration [see Eq. (\ref{coupling_j})] is rather sensitive to the details of the lattice (i.e., number of waveguides, length, etc). First of all, as shown in table \ref{tab1}, the optimal detunings 
are smaller or at most comparable to the couplings of the corner site to its neighbours. 
Furthermore, as shown in Fig. \ref{fig8} the corner site may acquire at least 10\% of the input power as the light propagates from $z=0$ to $L$, which is not a  negligible amount. These two observations together suggest that there 
is no solid ground for omission of the corner site relative to the others, and thus the derivation of an effective chain cannot be justified.
 
\begin{figure}
\centering
\includegraphics[scale=0.9]{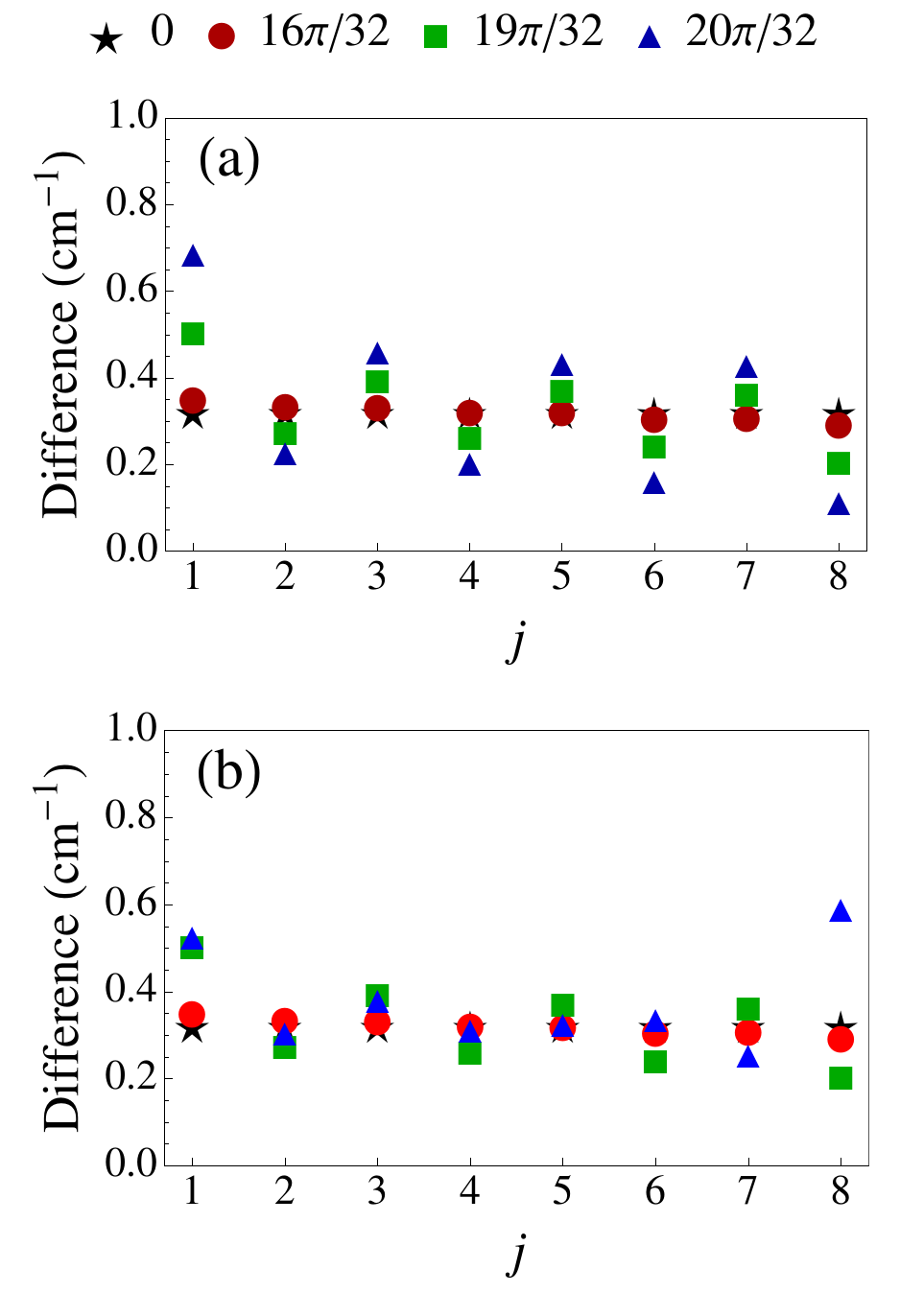}
\caption{(Color online) Spectrum of matrix ${\mathfrak J}$ for various bend angles. 
(a) Difference between successive eigenvalues before optimization of the corner site.  (b) As in (a),  after optimization of the corner site. Other parameters as in Fig. \ref{fig2} and table \ref{tab1}.
}
\label{fig9}
\end{figure}

The coupling configuration under consideration is a member of a large class of state-transfer 
Hamiltonians, whose operation relies on the commensurate eigenenergies \cite{ReviewsPST, EdVolume}. As explained in Ref. \cite{GenericBends}, for such a type of Hamiltonians a corner defect  minimizes the bending losses by  rearranging the spectrum that has been distorted by the bend. To confirm this once more, in Fig. \ref{fig9} we plot  the separation between successive eigenvalues of the matrix ${\mathfrak J}$ in Eq. (\ref{eom}), for different angles before and after optimization of the corner site. For the unbent chain ($\theta = 0$)  the eigenvalues are commensurate (i.e., equidistant), and that is why the coupling configuration of Eq. (\ref{coupling_j}), ensures ideally perfect transfer of light between the two outermost waveguides. As we bend the lattice, however, the commensurate nature of the eigenvalues is distorted, and the distortion is getting larger for sharper bends [see Fig. \ref{fig9}(a)].  As depicted in Fig. \ref{fig9}(b), the inclusion of a defect at the corner site of the bend tends to restore the relative position of the eigenvalues (i.e., the deviations from the case of the unbent chain are getting smaller). The remaining deviations at the borders are not of great importance since the contribution of eigenvectors with small/large indices to the evolution of the system is negligible \cite{GenericBends}.

\section{Concluding remarks}
\label{sec4}

We have analyzed the effects of bends on the transport of photonic signals between 
the two outermost waveguides of a  2D photonic lattice with engineered nearest-neighbour 
couplings that operates in the linear regime. In contrast to previous studies, in the present scheme the suppression of dispersion effects and the faithful transport of light  does not rely on Kerr nonlinearities, but rather on  the  engineering  of 
judicious couplings between nearest neighbours. It has been shown that our scheme works 
reliably for bends at least  up to $\theta_c = 90^\circ$ (with respect to the unbent chain). Sharper bends (with $\theta>\theta_c$)  have been shown to distort the transport, with the distortion being more pronounced for asymmetric waveguides. In this case, one has to find ways for minimizing bending effects, and in this direction it has been shown that the inclusion of a defect at the corner site can be a rather useful approach. Although our findings suggest that the present scheme outperforms its nonlinear (soliton-based) counterparts, further analysis is required for definite conclusions in this respect. 

Laser-written buried waveguides in glasses typically have elliptic shape, due to the beam focus, and they exhibit ``form 
birefringence", as a result of which fields with different polarizations experience different effective refractive indices \cite{FSTrealization}. 
Moreover, due to the formation of self-aligned nanogratings in the material during the irradiation process, one may also have ``material birefringence" \cite{San-etalPRl10,nano-grating}. 

Birefringence is a detrimental effect for quantum circuits that are intended for efficient guide and manipulation of qubits that are encoded in the polarization of photons. 
In general, the shape of the waveguides can be controlled efficiently by shaping the writing beam using standard 
techniques \cite{SymWGwriting}, and thus ``form birefringence" can be, in principle, eliminated. The elimination of ``material birefringence" 
is  also  feasible if one chooses the right material/substrate, and the right combination of writing  parameters  (i.e., wavelength, duration and energy of the laser pulses, repetition rate,
objective numerical aperture, translation speed, etc). In this way, the overall birefringence can be reduced 
by at least an order of magnitude facilitating thus the design of photonic primitives (e.g.,  directional couplers),  
that operate efficiently for polarization-encoded qubits \cite{San-etalPRl10}. 
One has to keep in mind, however, that  ``material birefringence" can be useful in the design of crucial polarization-sensitive components of quantum circuits, such as integrated wave plates \cite{Waveplates}, polarization routers \cite{nano-grating}, etc. 
In this context, for instance, one can have waveguides that allow for transmission of light with specific polarization, whereas light with the  orthogonal polarization is totally reflected.

For the sake of simplicity, the present analysis of bending losses  has been restricted to one of the lowest polarization modes (the calculations and conclusions for the other polarization are the same given that birefringence is not included in our formalism). Strictly speaking, the present results are valid for optical networks and communication schemes in  which qubits are not encoded in the polarization of the signal. In the case of polarization-based qubits,  the present 
results and conclusions are expected to be valid only for photonic lattices that are polarization-independent i.e., non-birefringent.  As mentioned above, currently available technologies allow for fabrication of such a type of lattices \cite{San-etalPRl10}. 

In closing it has  to be emphasized that the precise value of the critical angle $\theta_c$  beyond which bending effects cannot be ignored, depends on the details of particular set-up under consideration. The present analysis can be performed for any coupling configuration, and  it is pertinent to on-going  experiments on photonic lattices, in the framework of which  
a point-to-point link that relies on the coupling configuration discussed here has been 
demonstrated \cite{PSTrealization}. When combined with the ideas of \cite{3DPLs}, the present findings suggest that linear waveguide arrays with  engineered nearest-neighbour couplings can be used as building blocks of fundamental optical primitives that perform more demanding communication tasks such as routing, splitting, blocking, as well as logical functions.  

\section{Acknowledgements}
The author acknowledges with pleasure discussions at various times with M. Bellec on experimental  issues pertaining to laser-written photonic lattices. He is also  
grateful to P. Lambropoulos, T. Brougham and M. Bellec, for helpful comments on this paper.

\appendix

\begin{appendix} 
\section{Calculation of eigenmodes of a rectangular waveguide}
Here we discuss briefly an analytic solution of Eq. (\ref{Hy_eq}) using the separation of variables.  
This is a well known standard procedure and the details are discussed  in various standard textbooks and papers \cite{Marc69,book1,book2}. 
Setting  
\bea 
\label{HXY}
{\mathscr H}_x^{(j)}(x,y) = {\mathscr X}(x){\mathscr Y}(y),
\eea
Eq. (\ref{Hy_eq}) splits into two independent parts (one for each direction). The solutions  (up to normalization factors) are the 
following 
\bea
{\mathscr  X}(x) = \left \{
\begin{array}{ll} 
\cos(k_x x), &  |x| \leq \frac{\Delta x}2\\
\cos(k_x  \Delta x/2)e^{-\gamma_x(x-\Delta x/2)} &  |x| > \frac{\Delta x}2 
\end{array} \right. .
\label{Xpart}
\eea
and 
\bea
{\mathscr  Y}(y) = \left \{
\begin{array}{ll} 
\cos(k_y y), &  |y| \leq \frac{\Delta y}2\\
\cos(k_y  \Delta y/2)e^{-\gamma_y(y-\Delta y/2)}, &  |y| > \frac{\Delta y}2 
\end{array} \right. 
\label{Ypart}
\eea
with 
\bea
\label{gammax}
&&\gamma_x^2+k_x^2 = k^2(n_g^2-n_s^2)\\
\label{gammay}
&&\gamma_y^2+k_y^2 = k^2(n_g^2-n_s^2)\\
&&\beta^2 = k^2n_g^2-k_x^2-k_y^2.
\label{beta}
\eea
Boundary conditions on the electric field  imply also that 
\bea
\label{atanx}
&&k_x  \Delta x = \arctan\left ( \frac{\gamma_x}{k_x}\right )\\
&&k_y \Delta y = \arctan\left ( \frac{\gamma_y}{k_y}\right )
\label{atany}
\eea
where we have used the fact that $n_g\approx n_s$ \cite{remark2}.  
Equations (\ref{gammax}) - (\ref{atany}) form a closed set and determine all the parameters entering Eq. (\ref{HXY}). 

\end{appendix}


\begin{thebibliography}{99}
\bibitem{remark1} Discrete solitons are self-trapped states that exist as a result of 
balance between linear coupling effects and material Kerr nonlinearity. Their existence is 
only possible for input powers above a certain power threshold.
\bibitem{remark3} In the present model, there will be  some dependence on $\theta$ 
because of the approximation of the refractive index at the shaded (corner) areas in Fig. \ref{fig1}. 
For symmetric waveguides, however, this dependence is very weak.
\bibitem{remark4} 
One can always  engineer new coupling configurations that go beyond nearest neighbours, and ensure faithful transfer of signals between the two ends of the chain. The engineering of networks 
that perform certain tasks has been discussed extensively in the literature, and  goes beyond the scope of the present work \cite{EdVolume,HamEng}
\bibitem{remark5}
In order to ensure $\min\{r_{l,m}\}  > r_{C,C\pm 1}$ (or equivalently $\max\{{\mathscr J}_{l,m}\} < {\mathscr J}_{C,C\pm 1} $), we have restricted our simulations to  $\theta > 120^\circ$.
\bibitem{remark2} For the sake of simplicity, the above equations are given for 
a waveguide that is centred at $(x,y)=0$, and their generalization to arbitrary centres 
is straightforward.
\end{thebibliography}
\end{document}